\documentclass[letterpaper, 10 pt, conference]{ieeeconf}  

\IEEEoverridecommandlockouts                              

\usepackage{graphics} 
\usepackage{epsfig} 
\usepackage{times} 
\usepackage{amsmath} 
\usepackage{amssymb}  
\usepackage{algorithm}
\usepackage{algpseudocodex}
\usepackage[symbol]{footmisc}
\usepackage{hyperref}
\usepackage{xcolor}

\title{\LARGE \bf
Funnel-based Reward Shaping for Signal Temporal Logic Tasks in Reinforcement Learning
}

\author{Naman Saxena$^\dagger$, Gorantla Sandeep$^\dagger$, and Pushpak Jagtap
\thanks{This work was supported in part by the Google Research Grant, the ARTPARK, and the SERB Start-up Research Grant.}
\thanks{N. Saxena is with the Department of Computer Science and Automation, and G. Sandeep and P. Jagtap are with the Robert Bosch Center for Cyber-Physical Systems at the Indian Institute of Science, Bangalore, India. 
{\tt\small \{namansaxena,sgorantla,pushpak\}@iisc.ac.in}
}}

\begin{document}

\maketitle
\thispagestyle{empty}
\pagestyle{empty}

\begin{abstract}
Signal Temporal Logic (STL) is a powerful framework for describing the complex temporal and logical behaviour of the dynamical system. Numerous studies have attempted to employ reinforcement learning to learn a controller that enforces STL specifications; however, they have been unable to effectively tackle the challenges of ensuring robust satisfaction in continuous state space and maintaining tractability. In this paper, leveraging the concept of funnel functions, we propose a tractable reinforcement learning algorithm to learn a time-dependent policy for robust satisfaction of STL specification in continuous state space. We demonstrate the utility of our approach on several STL tasks using different environments.
\end{abstract}

\section{Introduction}
\footnotetext[2]{Authors contributed equally}
Temporal logic is an effective method of formally defining complex tasks involving spatial, temporal, and logical constraints \cite{baier2008principles}. The expressiveness of predicate logic combined with temporal dimension led to its widespread use of Linear Temporal Logic (LTL) \cite{1,2} in designing the specification of dynamical systems. Signal Temporal Logic (STL) \cite{maler2004monitoring} extends the applicability of LTL by allowing us to specify a behaviour for a fixed time interval. For example, a warehouse robot needs to reach particular locations in specific time intervals. The STL framework proves to be effective in efficiently capturing and modeling such requirements.

Several works in literature use the dynamics model to design controllers to enforce STL specifications (see \cite{9} and references therein). However, they are restricted to a limited class of systems and a fragment of specifications. Recently, controller synthesis for STL specification without a mathematical model of systems using Reinforcement Learning (RL) has started gaining attention. Not only is RL beneficial in the context of STL specifications, but at the same time, STL facilitates the designing of rewards for RL in a structured manner, avoiding the loopholes in developing rewards in a heuristic manner \cite{2}. \cite{3} used Q-learning \cite{10} to achieve tasks defined using STL by maximizing the robustness of STL satisfaction. \cite{3} defined the $\tau-$MDP framework to allow each state to store the history of states. Storing the history of states is required to check for the satisfaction of STL formulas and, at the same time, raises doubts about the tractability of the proposed method. Further, \cite{4} tries to solve the tractability issue of Q-learning by proposing the use of flag variables. Flag variables are used to avoid storing the history of states and check the satisfaction of STL formulas. One issue with this work is that it does not consider robustness values. \cite{5} proposed a solution for multi-agent system using Deep Q-learning algorithm \cite{11}. The authors used Deep Q-Network (DQN) to overcome the state-space explosion in the multi-agent setting. Their approach considers robustness value for the STL formula but again suffers from the drawback of storing the history of states.

On the other hand, a funnel-based control approach is employed to enforce the satisfaction of a fragment of STL specifications \cite{6} by formulating exponentially decaying constraint functions referred to as 'funnels'. The authors in \cite{6} introduce a continuous and closed-form controller construction for enforcing those fragments of STL specifications. Achieving this objective necessitates making certain assumptions about the systems (such as requirements of control-affine systems, fully/overactuated systems, and unbounded inputs) and about specifications (such as STL tasks excluding 'OR' logical operators, convex predicates, and logical operations over temporal operators). Nevertheless, this approach holds great promise in effectively separating the temporal and logical components within an STL specification while capturing robustness on satisfaction.    
  
Leveraging the advantages of the results in \cite{6}, this work proposes a funnel-based approach for reward shaping in reinforcement learning algorithms to enforce STL specification in a tractable manner. We demonstrated that the proposed approach resolves the issue of tractability by eliminating the requirement of storing state history (\cite{3}, \cite{5}, \cite{wang2023multi}, \cite{ikemoto2022deep}). This key enhancement enables the extension of our proposed approach to continuous-state environments. Furthermore, we present evidence that, for the first time, the proposed approach integrates robustness considerations while enforcing STL specifications in the RL framework in the context of continuous-state spaces. Moreover, by leveraging a learning framework, we can relax several assumptions made in \cite{6} concerning systems and specifications. For instance, we observe that our approach can handle the logical operator `OR', any type of predicate, and conjunction between temporal operators while allowing us to provide results for any general nonlinear systems with input constraints. It is important to note that the advantages we observe are based on empirical findings and lack formal guarantees, unlike the methodology presented in \cite{6}. Finally, we demonstrated the effectiveness of the proposed approach with several simulation results on different environments and real-world sim-to-real transfer. 

\section{Preliminaries}\label{preliminaries}

\subsection{Deep Q-learning}\label{sc:2.1}
Reinforcement learning (RL) is a learning paradigm based on the framework of Markov Decision Processes (MDP) \cite{15}. An MDP is defined by a tuple $\mathcal{M}=(\mathcal{S}, \mathcal{A},r, \mathcal{P}, \pi, \gamma)$, where $\mathcal{S} \subset \mathbb{R}^m $ refers to the continuous state space, $\mathcal{A}$ refers to the discrete action space and $r:\mathcal{S}\times\mathcal{A} \mapsto \mathbb{R}$ is the reward function. Further, $\mathcal{P}(\cdot|s,a)$ is the transition probability function defined as $\mathcal{P}:\mathcal{S}\times\mathcal{A} \mapsto \mu(\cdot)$. $\mu: \mathcal{B}(\mathcal{S}) \mapsto [0,1]$ is a probability measure and $\mathcal{B}(\mathcal{S})$ is the Borel $\sigma-$algebra on state space $\mathcal{S}$. Here, $\pi: \mathcal{S} \mapsto \Delta(\mathcal{A})$ is a stochastic policy (controller) and $\Delta(\mathcal{A})$ is the probability simplex over action space $\mathcal{A}$. $\gamma \in (0,1)$ is the discount factor. The policy $\pi$ is obtained by optimizing the long-term discounted reward objective function $\eta(\pi)$ as defined below:
\begin{equation}\label{eq:6}
\eta(\pi) = E\Big[\sum_{t=0}^{\infty}\gamma^{t}r(s_t, a_t) \Big],
\end{equation}
where $s_t$, $a_t$ denotes the state and action taken at time $t$. To solve the above optimization problem, Q-learning \cite{10} is one of the most widely used RL algorithms. It uses $\epsilon-$greedy policy \eqref{eq:8} based on the Q-value function \eqref{eq:7} to explore and optimize the objective function in \eqref{eq:6}. 
\begin{align}
\hspace{-0.5em}Q^{\pi}(s_t, a_t) 
&= E\Big[\sum_{k=t}^{\infty}\gamma^{k-t}r(s_k, a_k)|s_t,a_t\Big]\nonumber\\ &= E\Big[r(s_t, a_t) + \gamma Q^{\pi}(s_{t+1},a_{t+1})|s_t,a_t\Big]. \label{eq:7}\\
\pi(a|s) &= 
{
\begin{cases}
1 - \epsilon + \dfrac{\epsilon}{|\mathcal{A}|} & a = \arg\max_{a'} Q^{\pi}(s,a') \\
\dfrac{\epsilon}{|\mathcal{A}|} & \text{otherwise}. 
\end{cases}
}\label{eq:8}
\end{align}
Here, $Q^{\pi}(s,a)$ is the Q-value function \eqref{eq:7} for $(s,a) \in \mathcal{S}\times\mathcal{A}$ pair that denotes the long-term discounted reward achieved after taking action $a$ in state $s$ and following the policy $\pi$ after that. 
The Q-learning finds optimal policy $\pi^{*}$ by finding solution to Bellman equation 
\begin{equation}\label{eq:9}
Q^{\pi^{*}}(s, a) = E[ r(s, a) + \gamma \max_{\bar{a} \in \mathcal{A}} Q^{\pi^{*}}(s',\bar{a})]     
\end{equation}
that ensures $Q^{\pi^{*}}(s,a) \geq Q^{\pi}(s,a)$ $\forall \pi$, $s\in\mathcal{S}$, and $a\in\mathcal{A}$. 

Deep Q-learning \cite{11} is a function approximation-based Q-learning algorithm that uses a neural network to learn optimal policy online using a replay buffer. The algorithm uses a neural network with parameters $\theta$ to approximate Q-value function $Q_{\theta}(s,a)$ that satisfy the Bellman equation \eqref{eq:9}. 

\subsection{Signal Temporal Logic}\label{sc:2.2}
Signal temporal logic (STL) \cite{maler2004monitoring} provides a formal framework to capture high-level specifications containing spatial, temporal, and logical constraints. It consists of a set of predicates $\varphi$ that are evaluated using predicate functions $h:\mathcal{S} \rightarrow \mathbb{R}$ as $\varphi:=\left\{ \begin{array}{l} \text{True, \ \ \ \ \ if } h(s) \geq 0 \\ \text{False, \ \ \ \ if } h(s) < 0 \end{array} \right.$.
The syntax for an STL formula $\phi$ is given by:
\begin{align*}
    \phi::=\text{True} \mid \varphi \mid \neg\phi \mid \phi_1\wedge\phi_2 \mid \phi_1\vee\phi_2 \mid F_{[a, b]}\phi \mid G_{[a, b]}\phi,
\end{align*}
where $a, b \in \mathbb{R}^+_0$ with $a\leq b$, $\phi_1$ and $\phi_2$ are STL formulas, $\neg$, $\land$ and $\lor$ are logical negation, conjunction and disjunction operator, respectively; and $F$ and $G$ are temporal \textit{eventually} and \textit{always} operators, respectively. The relation $s_t\models \phi$ indicates that the signal $s:\mathbb{R}_{\geq 0} \mapsto \mathcal{S}$ satisfies the STL formula $\phi$ at time $t$. The STL semantics for a signal $s$ is recursively defined as follows:
\begin{alignat}{2}\label{eq:1}
    &s_t \models \varphi && \Longleftrightarrow \varphi \;\;\text{is True} \nonumber\\
    &s_t \models \neg \phi && \Longleftrightarrow \neg(s_t\models \phi)\nonumber\\
    &s_t \models \phi_1 \wedge \phi_2 && \Longleftrightarrow s_t \models \phi_1 \wedge s_t \models \phi_2 \nonumber\\
    &s_t \models \phi_1 \vee \phi_2 &&\Longleftrightarrow s_t \models \phi_1 \vee s_t \models \phi_2 \nonumber\\
    &s_t \models F_{[a, b]}\phi && \Longleftrightarrow \exists t' \in \left[t+a, t+b\right]\text{ s.t. }s_{t'} \models \phi \nonumber\\
    &s_t \models G_{[a, b]}\phi &&\Longleftrightarrow \forall t' \in \left[t+a, t+b\right] \text{ s.t. } s_{t'}\models\phi.
\end{alignat}
Next, we recall the robust semantics for STL formulas introduced by \cite{donze2010robust}, which will later be used to construct rewards.
\begin{equation}
\begin{split}
\rho_{\varphi}(s_t) &= h(s_t) \\
\rho_{\neg\phi}(s_t) &= -\rho_\phi(s_t) \\
\rho_{\phi_1 \land \phi_2}(s_t) &= \min(\rho_{\phi_1}(s_t), \rho_{\phi_2}(s_t))\\
\rho_{\phi_1 \lor \phi_2}(s_t) &= \max(\rho_{\phi_1}(s_t), \rho_{\phi_2}(s_t))\\
\rho_{F_{[a,b]}\phi}(s_t) &= \max_{t'\in [t+a,t+b]}\rho_\phi(s_{t'})\\
\rho_{G_{[a,b]}\phi}(s_t) &= \min_{t'\in [t+a,t+b]}\rho_\phi(s_{t'}).
\end{split}
\end{equation}
In this paper, we consider the following fragment of STL: 
\begin{equation}\label{eq:2}
\begin{split}
    \psi &: = \varphi \;\;|\;\;\neg \psi \;\;|\;\; \psi_1 \land \psi_2\;\;|\;\; \psi_1 \lor \psi_2 \\
    \phi_{[a,b]} &: =  F_{[a,b]}\psi \;\;|\;\; G_{[a,b]}\psi \mid F_{[a,c_1]} G_{[c_2,b]}\psi\\
    \Phi &:= \bigwedge_{i=1}^k \phi_{[a_i,b_i]} \;\;|\;\; \bigwedge_{i=1}^k \phi_{[\alpha_i,\beta_i]},
\end{split}    
\end{equation}
where $0\leq a\leq c_1$, $ c_2\leq b$, $b_i<a_{i+1}$, $\forall i\in\{1,\ldots,k-1\}$, $\psi$ and $\phi$ denote non-temporal and temporal formulas, respectively. Further, there may exist temporal formula with overlapping time intervals such that for some $i,j\in\{1,\ldots,k-1\}$, $\alpha_i < \beta_j$ and $\alpha_j < \beta_i$.\\
In the next section, we discuss a funnel-based construction of rewards for deep Q-learning to learn a control policy enforcing the fragment of STL specifications given in \eqref{eq:2}. 




\section{Proposed Approach}\label{Proposed_Approach}
In this section, we describe the utilization of funnel-based control concepts to construct time-varying rewards capturing robust satisfaction of STL specifications.

\subsection{Construction of Rewards using Funnel Functions}\label{funnel_reward}
Funnel based approach was first used by \cite{6} to develop controllers that satisfy a fragment of STL specifications for known control systems. Several works in the literature now build upon this direction \cite{7,14}. \cite{6} 
proposed to find a controller that satisfies the following relation:
\begin{equation}\label{eq:5}
\forall \; t\geq0,\;\; -\gamma(t) + \rho_{max} < \rho_{\psi}(s_t) < \rho_{max}, 
\end{equation}
where $\gamma(t)$ is a non-increasing and continuously differentiable positive function referred to as \textit{funnel} and defined as $\gamma(t)= (\gamma_{0}-\gamma_\infty)\mathsf{e}^{-lt}+\gamma_\infty$, where $\gamma_{0},\gamma_\infty$, and $l$ are positive constants with $\gamma_0\geq\gamma_\infty$, and $\rho_{max}$ is the maximum robustness defined for the system for corresponding non-temporal specification $\psi$ and obtained as $\rho_{max}=\max_{s\in\mathcal{S}}\rho_\psi(s)$.

Let us consider the STL fragment defined in \eqref{eq:2}. The STL formula $\Phi$ can consist of a single eventually $(F)$ or always $(G)$ operator, or it could contain these operators combined using a conjunction operator. 
The parameter $l$ of funnel function $\gamma(t)$ for $F_{[a,b]}\;\psi$, $G_{[a,b]}\;\psi$, and $F_{[a,c_1]} G_{[c_2,b]}\psi$ is chosen as given in Table I, while $\gamma_0 = \rho_{max} - \min_{s\in\mathcal{S}}\rho_{\psi}(s)$ and $\gamma_\infty\in(0,\min(\gamma_0,\rho_{max}))$ for all the temporal operators.

  \begin{table}[h!]
   \centering
    \begin{tabular}{l|lll}
           & $t^*$ &  $l$ \\ \cline{1-3}
     $G_{[a,b]}\psi$   &  $t^*=a$   & $\frac{1}{t^*}\ln\frac{\gamma_0-\gamma_{\infty}}{\rho_{max}-\gamma_{\infty}}$ \\
     $F_{[a,b]}\psi$ & $t^*\in[a,b]$   & $\frac{1}{t^*}\ln\frac{\gamma_0-\gamma_{\infty}}{\rho_{max}-\gamma_{\infty}}$\\
     $F_{[a,c_1]} G_{[c_2,b]}\psi$ &$t^*\in[a+c_2, c_1+c_2]$ & $\frac{1}{t^*}\ln\frac{\gamma_0-\gamma_{\infty}}{\rho_{max}-\gamma_{\infty}}$\\
    \end{tabular}\label{tb:1}
    \caption{Selection of funnel function parameter $l$.}
    \end{table}

\begin{figure}
    \centering
    \includegraphics[height = 4.3cm]{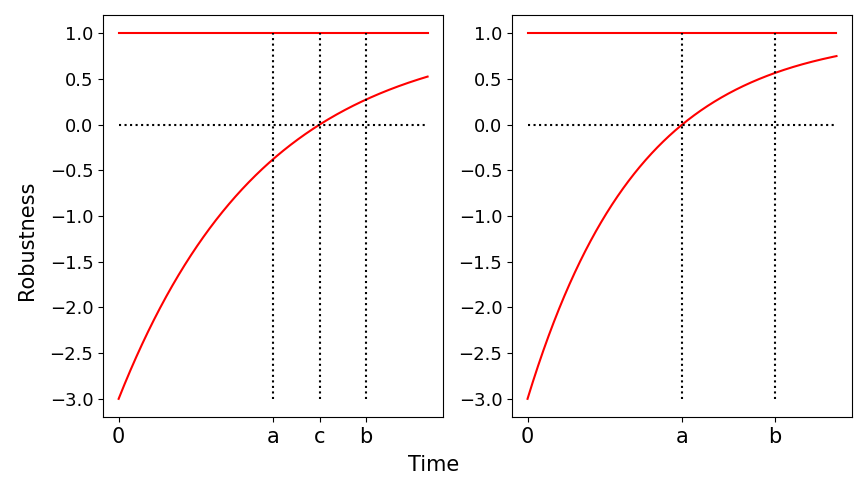}
    \vspace{-0.8em}
\caption{The funnel for eventually operator with $t^* = c \in [a,b]$ (left) and funnel for always operator (right).}
\label{fig:1}
\end{figure}

The illustration of the funnel for eventually and always operators is shown in Figure \ref{fig:1}.     
 The value for $l$ is chosen according to the interval for temporal operators. For $F_{[a,b]}$ operator $l$ is $\frac{ln((\gamma_0-\gamma_{\infty})/(\rho_{max}-\gamma_{\infty}))}{t^*}$, where $t^* \in [a,b]$, so that $\gamma(t^*) = 0$. Note that $t^*$ lies in $[a,b]$ because for the $eventually$ operator, we want the robustness to be positive at least once in the interval $[a,b]$. For $G_{[a,b]}$ operator $l$ is $\frac{ln((\gamma_0-\gamma_{\infty})/(\rho_{max}-\gamma_{\infty}))}{t^*}$, where $t^*=a$, so that $\gamma(a) = 0$ and the robustness is positive throughout the interval $[a,b]$. Similar reasoning follows for $F_{[a,c_1]} G_{[c_2,b]}$ operator. 

\begin{figure}
\centering
\includegraphics[height=4.3cm]{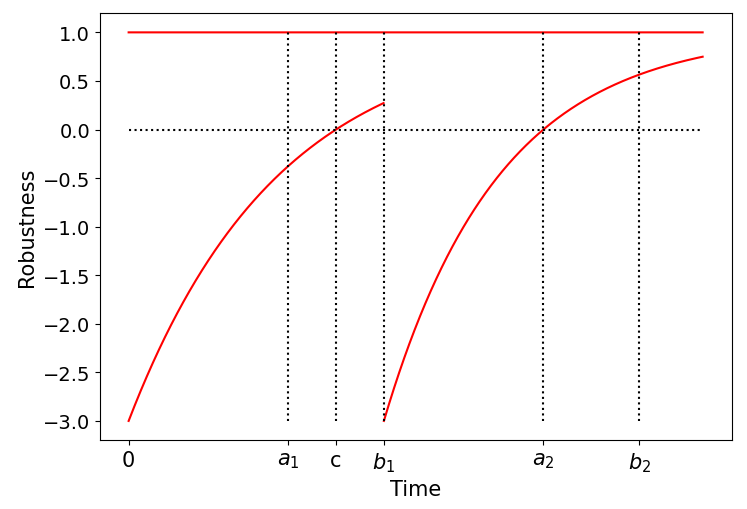}
\vspace{-0.9em}
\caption{The funnel function for conjunction of $F_{[a_1,b_1]}$ and $G_{[a_2,b_2]}$ operator.}
\label{fig:2}
\end{figure}

Now we will discuss cases where temporal operators appear in conjunction. Let us take the STL formula $\Phi = F_{[a_1,b_1]}\;\psi_1 \land G_{[a_2,b_2]}\;\psi_2 $ with $b_1<a_2$, where $\psi_1$ and $\psi_2$ are as defined in \eqref{eq:2}. The funnel function for $\Phi$ is given as
\begin{equation}\label{eq:11}
\resizebox{\linewidth}{!}{$%
 \gamma(t) = 
 {
 \begin{cases}
(\gamma_{0}-\gamma_{\infty})\mathsf{e}^{-\ln(\frac{\gamma_{0}-\gamma_{\infty}}{\rho_{max}-\gamma_{\infty}})\frac{t}{c}} + \gamma_{\infty},   & \text{for } 0\leq t \leq b_1,\\
(\gamma_{0}-\gamma_{\infty})\mathsf{e}^{-\ln(\frac{\gamma_{0}-\gamma_{\infty}}{\rho_{max}-\gamma_{\infty}})\frac{t-b_1}{a_2-b_1}} +\gamma_{\infty},   & \text{for } t > b_1,\\
 \end{cases}
 }$%
 }
\end{equation}
and the plot is shown in Figure \ref{fig:2}. In \eqref{eq:11}, $t^* = c$ lies in $[a_1,b_1]$. For $t\leq b_1$, the funnel function $\gamma(t)$ is defined according to $F_{[a_1,b_1]}$ operator and for $t > b_1$, the funnel function is defined according to $G_{[a_2,b_2]}$ operator. In this case, the $F$ operator is considered with $G$, but the funnel function can also be designed similarly to handle two $F$ operators and/or two $G$ operators. Further, this method of designing funnel function is not limited to two operators but can be extended to several operators in conjunction.

Given the construction of funnel function $\gamma(t)$ for the temporal part and the robustness measure $\rho_{\psi}(s_t)$ for the non-temporal part of the STL formula, we can now define the reward function for the deep Q-learning algorithm as
\begin{equation}\label{eq:12}
r'(s_t, a_t, t) = \rho_{\psi}(s_t) + \gamma(t) - \rho_{max}.
\end{equation}
The reward function in \eqref{eq:12} is positive at time $t$ if the current state of the system (agent) follows the bounds given in \eqref{eq:5}; otherwise, it is negative. Further, the reward is more positive if $\rho_{\psi}(s_t)$ is close to $\rho_{max}$ and more negative as it is farther away from the lower bound in \eqref{eq:5} in the negative direction. 

In general, one needs a history of states to check whether a predicate is satisfied for the time interval of the temporal operator. However, in the case of a funnel-based reward, the temporal satisfaction is captured using time-varying funnel constraints as discussed above. Thus, being inside the funnel constraints at each time instance indicates that the predicate is satisfied for the given time interval. Hence, we do not need to explicitly store the history of the states.
\subsection{Temporal Operator with Overlapping time intervals}
Let us consider the STL specification given in (\ref{eq:23}) using two temporal operator with overlapping time intervals. Here, in the specification $\alpha_1 < \beta_2$ and $\alpha_2 < \beta_1$.
\begin{equation}\label{eq:23}
 \Phi = \underbrace{G_{[\alpha_1,\beta_1]}\phi_1}_{\psi_1} \land \underbrace{F_{[\alpha_2,\beta_2]}\phi_2}_{\psi_2}    
\end{equation}
Let $r'(s_t, a_t, t, \psi_1)= \rho_{\psi_1}(s_t) + \gamma_1(t) - \rho_{max,1}$ and $r'(s_t, a_t, t, \psi_2)= \rho_{\psi_2}(s_t) + \gamma_2(t) - \rho_{max,2}$. Note that $r'(s_t, a_t, t, \psi)$ denotes the reward for predicate $\psi$. We design the reward by defining the reward as earlier in the time intervals with no overlap and taking the minimum of the reward for different predicates with overlapping time intervals. One can write a reward for the STL formula \eqref{eq:23} as:
\begin{equation*}
\resizebox{\linewidth}{!}{$%
\begin{split}
 &r'(s_t,a_t,t)\\ =& {
 \begin{cases}
r'(s_t, a_t, t, \psi_1), \hspace*{\fill} t \in [\alpha_1,\alpha_2)\\
\min(r'(s_t, a_t, t, \psi_1),r'(s_t, a_t, t, \psi_2)), \hspace*{\fill} t \in [\alpha_2, \min(\beta_1,\beta_2))\\
r'(s_t, a_t, t, \psi_2), \hspace*{\fill} t \in [\min(\beta_1,\beta_2), \max(\beta_1,\beta_2)]\\
 \end{cases}
 }
\end{split}$%
}
\end{equation*}
This approach is not limited to handling simultaneous overlapping between time intervals of two temporal operators but can also be used for an arbitrary number of simultaneously overlapping temporal operators (Algorithm \ref{alg:0}). Further, in Section \ref{sc:4.2}, we show experimentally that the reward structure proposed for overlapping intervals satisfies the STL specification robustly. 

\begin{algorithm}
\caption{Reward Calculation for Overlapping Interval}
\label{alg:0}
\begin{algorithmic}[1] 
\Function{Reward}{$s_t, a_t, t$}
\State intervals = $\{[a_i,b_i]\}_{i=0}^{n}$ \algorithmiccomment{$[a_i,b_i]$ is the time interval of $i^{th}$ temporal operator}
\State reward = 0 
\For{$i \in \{0,\cdots , n\}$}
\If{$t \in$ interval[i]}
\State reward = $\min$(reward , r'($s_t,a_t,t,\psi_i$))
\EndIf
\EndFor
\EndFunction
\Return reward
\end{algorithmic}
\end{algorithm}

\subsection{Time-aware Deep Q-learning}
Our time-aware Deep Q-learning algorithm uses the funnel-based time-dependent reward function in \eqref{eq:12}, which is not only a function of state and action but also a function of time $t$. The modified MDP is defined as $\mathcal{M}'= \{\mathcal{S},\mathcal{A}, r',\mathcal{P}, \pi',\gamma\}$, where $r': \mathcal{S}\times\mathcal{A}\times \mathbb{N} \cup \{0\} \mapsto \mathbb{R}$ is the new reward function. The stochastic policy is now defined as $\pi': \mathcal{S}\times \mathbb{N} \cup \{0\} \mapsto \Delta(\mathcal{A})$. The Markov property of the transition probability function is intact because the transition to $s_{t+1}$ still depends on $(s_{t},a_{t})$ and $a_t$ depends on $s_t$ and the current time $t$. Now, since the reward depends on time and consequently Q-value function also depends on time and is defined as follows: 
\begin{equation*}
\resizebox{\linewidth}{!}{$%
\begin{split}
Q^{\pi}(s_t, a_t, t) &= E\Big[\sum_{k=t}^{\infty}\gamma^{k-t}r(s_k, a_k,k)|s_t,a_t,t\Big] \\ &= E\Big[r(s_t, a_t, t) + \gamma Q^{\pi}(s_{t+1},a_{t+1},t+1)|s_t,a_t,t\Big].     
\end{split}$%
}
\end{equation*}
 Further, the policy will depend on time because we use the $\epsilon$-greedy policy, which depends on the Q-value function \eqref{eq:8}. The proposed method is summarized in Algorithm \ref{alg:1}. The next section discusses the results obtained using our time-aware Deep Q-learning algorithm.  

\begin{algorithm}[t]
\caption{Time-aware Deep Q-learning}
\label{alg:1}
\text{Initialize Q-value function parameter $\theta$.}\\ 
\text{Initialize target Q-value function parameter $\overline{\theta} \leftarrow \theta$}
\text{$\alpha$ is the step size for parameter update.}
\begin{algorithmic}[1] 
\State $k=0$, $s_{0}$ = env.reset(),$t=0$
\While{$k \leq$ total steps}
\State $a_t \sim \pi(\cdot|s_{t},t)$ \algorithmiccomment{$\pi$ is $\epsilon$-greedy policy }
\State $s_{t+1} \sim P(\cdot | s_t, a_t)$ and $r_t = r'(s_{t},a_{t}, t)$
\State Store $\{s_{t},a_{t}, r_{t}, s_{t+1}, t\}$ in Replay Buffer \algorithmiccomment{$s_{t+1} = s_{t}'$}
\If { $k\;\%\;eval\_freq == 0$}
\State Evaluate(agent)
\EndIf
\State Sample $\mathbb{B}_k = \{s_{i}, a_{i}, r_{i}, s_{i}', t_{i}\}_{i=0}^{M-1}$ from the Replay Buffer
\State Update $\;\theta \leftarrow \theta + \alpha\nabla_{\theta}\Big(\frac{1}{M}\sum_{i=0}^{M-1}\big(r'(s_{i},a_{i},t_i) + \gamma \max_{\bar{a}}Q_{\overline{\theta}}(s_{i}',\bar{a},t_i+1) - Q_{\theta}(s_{i},a_{i},t_i)\big)^{2}\Big)$
\If { $k\;\%\;target\_update\_freq == 0$}
\State Update $\;\;\overline{\theta} \leftarrow \theta$ 
\EndIf
\State $k=k+1$
\If {$s_{t+1}$ is terminal}
\State $s_{t} = $ env.reset(), $t=0$
\Else
\State $s_{t} = s_{t+1}$, $t=t+1$ 
\EndIf
\EndWhile

\end{algorithmic}
\end{algorithm}

\section{Experimental Results}\label{Experimental_Results}
We performed experiments on three case studies with various systems and STL properties to demonstrate the merits of the proposed approach. We trained the RL agent using a time-aware Q-learning algorithm (Algorithm \ref{alg:1}). 
\begin{figure*}[t]
    \centering
    \includegraphics[scale=0.41]{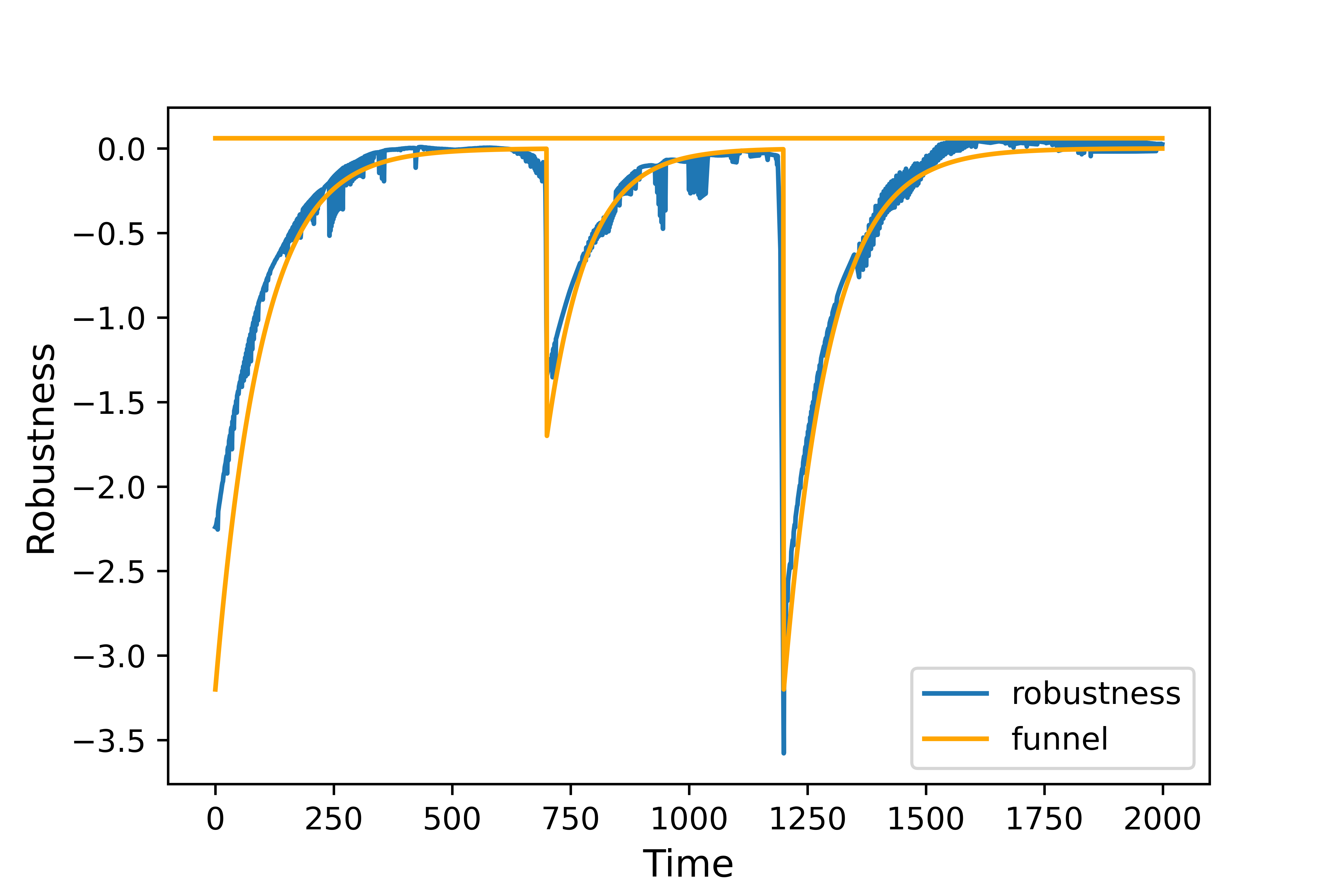}\hspace{-0.62cm}
    \includegraphics[scale=0.41]{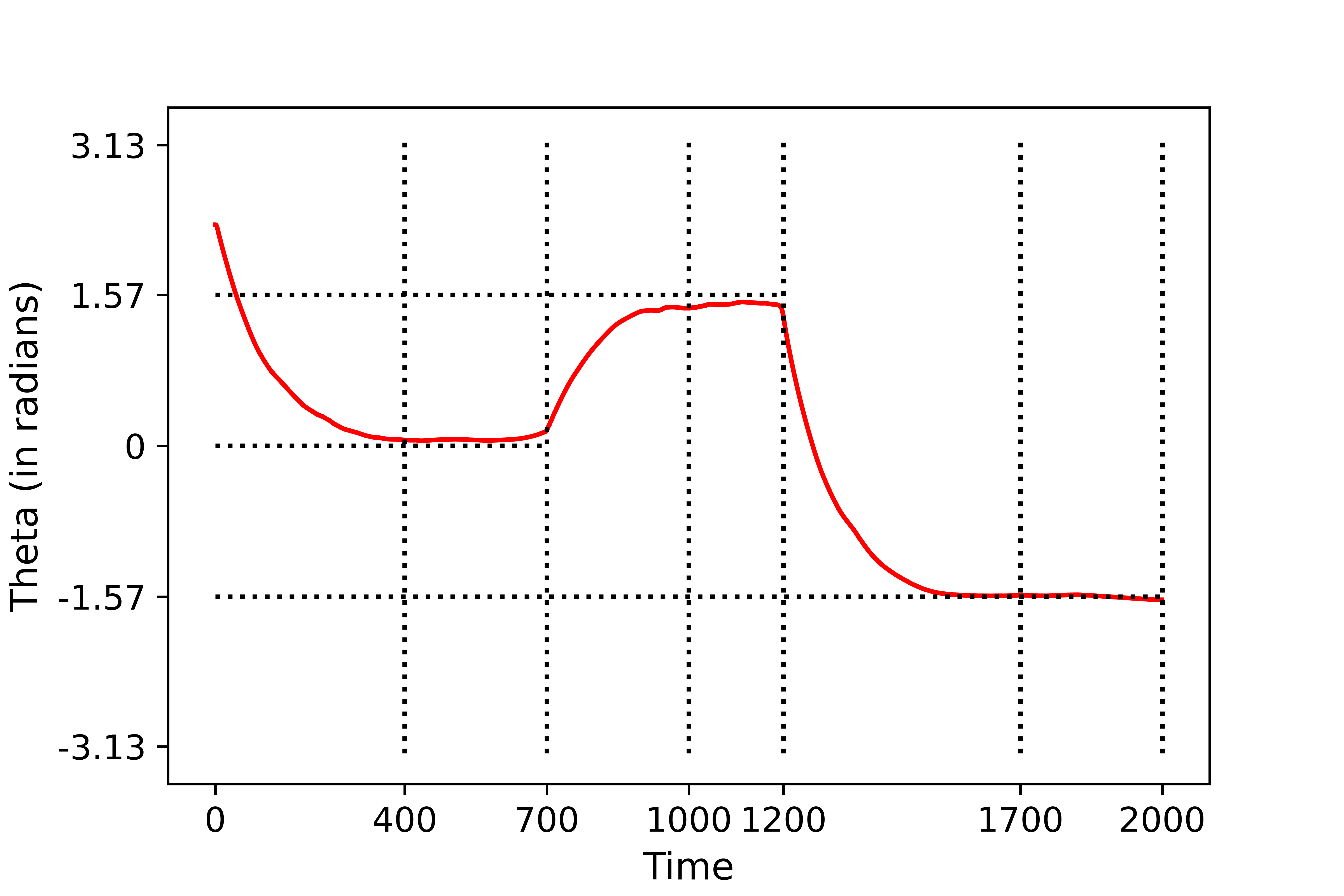}\hspace{-0.62cm}
    \includegraphics[scale=0.41]{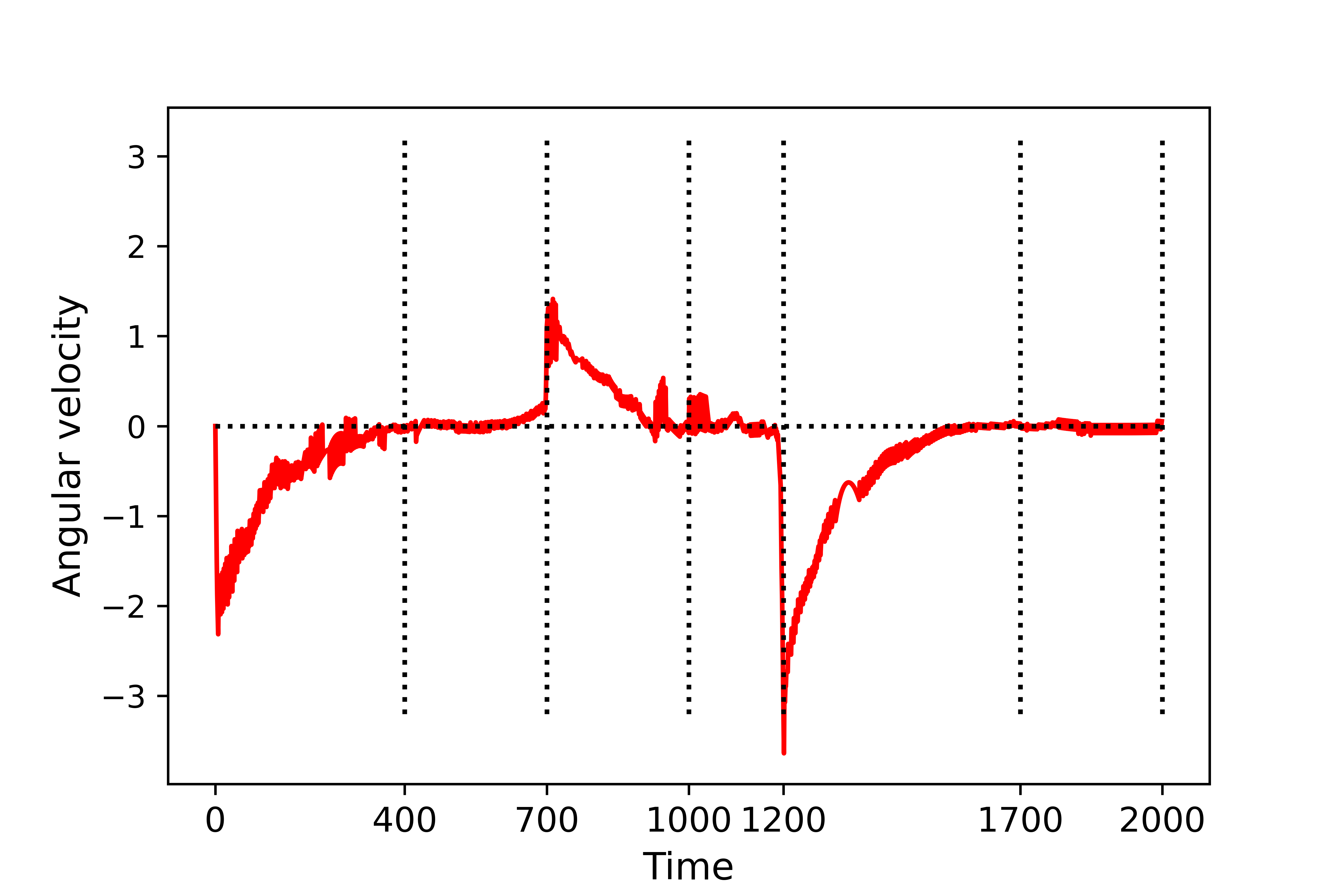}
    \vspace{-0.8em}
\caption{The evolution of robustness values (left), angle (middle), and angular velocity (right) of the pendulum.}
\label{fig:3}
\end{figure*}
\begin{figure*}[t]
    \centering
    \includegraphics[scale=0.41]{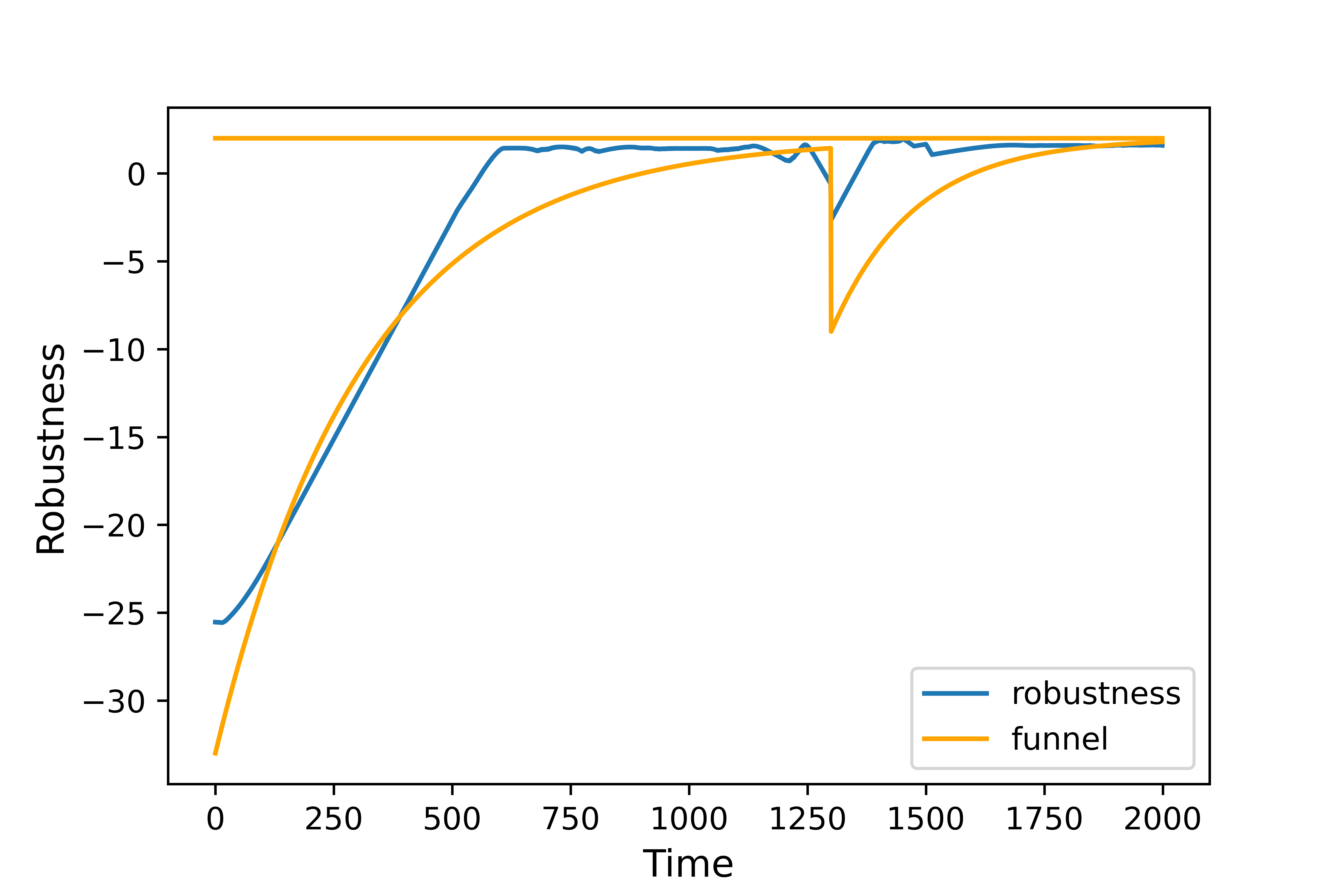}\hspace{-0.62cm}
    \includegraphics[scale=0.41]{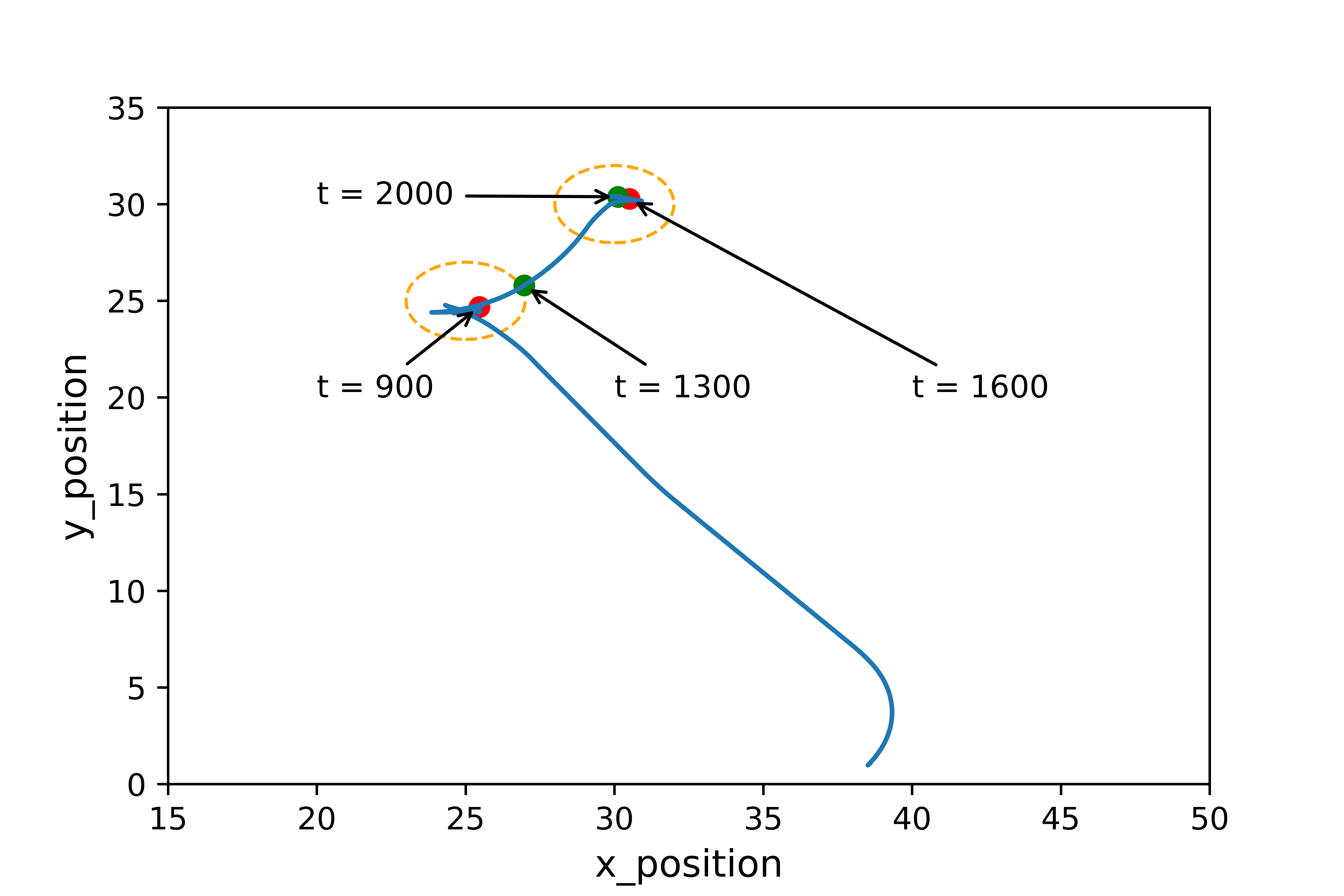}\hspace{-0.62cm}
    \includegraphics[scale=0.41]{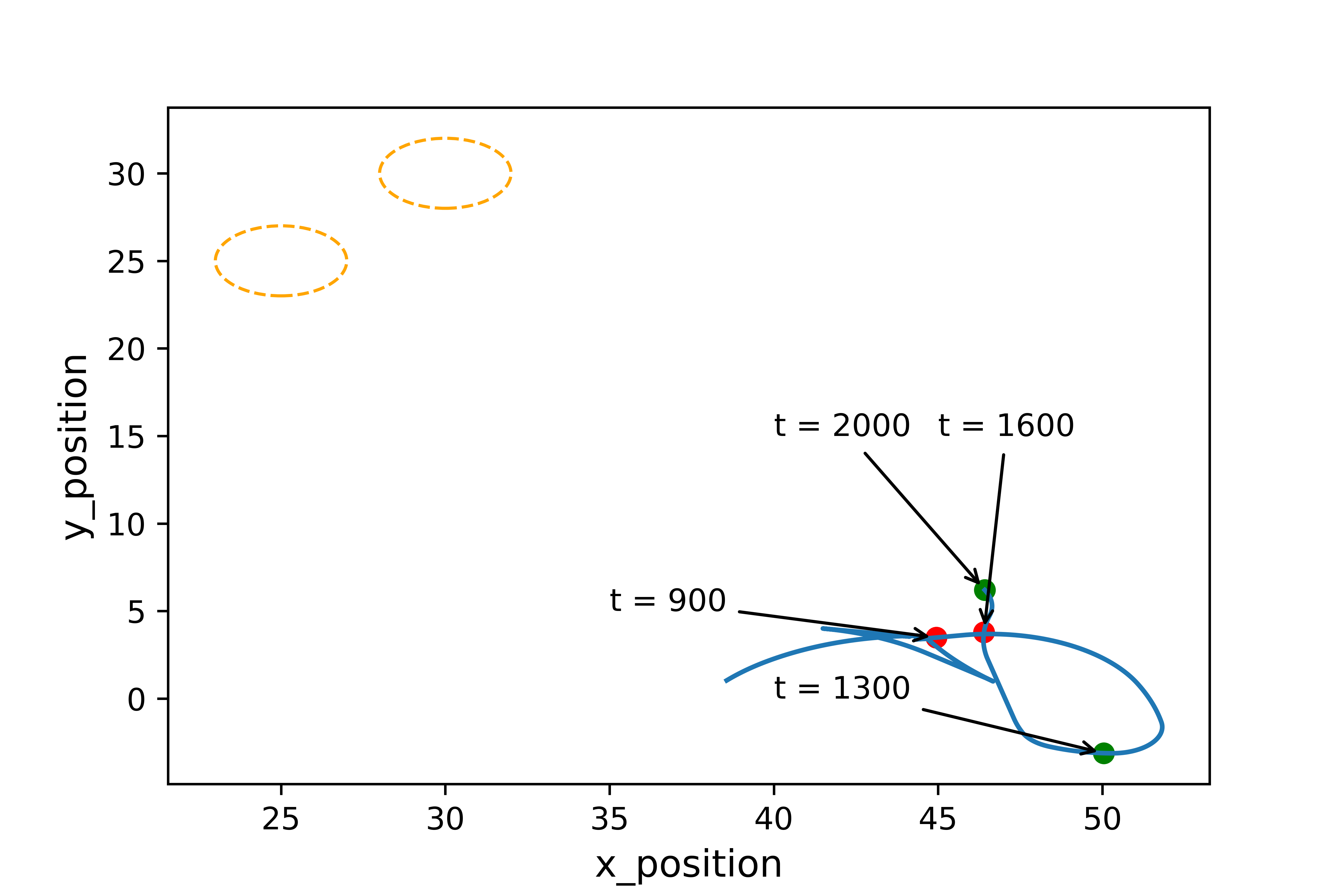}
\caption{The evolution of robustness values (left), the trajectory in the 2-d plane with controller learned using reward function with (middle) and without (right) funnel for a mobile robot.}
\label{fig:4}
\end{figure*}
\subsection{Pendulum System}
Consider an inverted pendulum model defined as:
\begin{align*}
    {\theta}_{t+1} &= \theta_t + \tau \omega_t,\\
    {\omega}_{t+1} &= \omega_t + \tau(\frac{g}{l}\sin{\theta_t} - \frac{\mu}{ml^2} \omega_t+  \frac{1}{ml^2}a_t),
\end{align*} 
where $\theta$, $\omega$, and $a\in\{-3,-2.9,\ldots,2.9,3\}$ are the angle of the pendulum, angular velocity, and actions representing torque applied, respectively. $\tau=0.01$ is the sampling time. The constants $g=9.8m/s^2, m=0.15, l=0.5m$, and $\mu=0.05$ represent acceleration due to gravity, mass, length of pendulum and friction coefficient, respectively. We consider the following STL specification:
\begin{equation}\label{eq:14}
\begin{split}
\Phi=&G_{[400,700]}\underbrace{(\lvert\theta\rvert\leq 0.05\land \lvert\omega\rvert \leq 0.05)}_{\psi_1} \\ &\land G_{[1000,1200]}\underbrace{(\lvert 1.57-\theta\rvert\leq 0.05\land \lvert\omega\rvert \leq 0.05)}_{\psi_2}\\ &\land G_{[1700,2000]}\underbrace{(\lvert-1.57-\theta\rvert\leq 0.05\land \lvert\omega\rvert \leq 0.05)}_{\psi_3}. 
\end{split}
\end{equation}
In simple words, the specification in \eqref{eq:14} says that "the pendulum should maintain $\theta=0$ and $\omega=0$ in the interval of 400 to 700 timesteps (which is equivalent to 4 to 7 seconds as $\tau$ is $0.01$ seconds) with a tolerance value of $0.05$. Subsequently, in the interval of 1000 to 1200 time steps, the pendulum should be balanced at $\theta=1.57$ and $\omega=0$ followed by $\theta=-1.57$ and $\omega=0$ from 1700 to 2000 time steps with a tolerance value of $0.05$". The funnel-based reward function for \eqref{eq:14} with $s_t=[\theta_t,\omega_t]$ is computed as discussed in Section \ref{funnel_reward} and given as follows:
\begin{equation}\label{eq:15}
\resizebox{\linewidth}{!}{$%
\begin{split}
 &r'(s_t,a_t,t)\\ =& {
 \begin{cases}
\rho_{\psi_1}(s_t) + (\gamma_{0,1}-\gamma_{\infty,1})\mathsf{e}^{-l_1 t}+\gamma_{\infty,1} - \rho_{max,1} & t \in [0,700]\\
\rho_{\psi_2}(s_t) + (\gamma_{0,2}-\gamma_{\infty,2})\mathsf{e}^{-l_2 (t-700)}+\gamma_{\infty,2} - \rho_{max,2} & t \in [700, 1200]\\
\rho_{\psi_2}(s_t) + (\gamma_{0,2}-\gamma_{\infty,3})\mathsf{e}^{-l_3 (t-1200)}+\gamma_{\infty,3} - \rho_{max,3} &  t \in [1700, 2000],\\
 \end{cases}
 }
\end{split}$%
}
\end{equation}
where $\rho_{\psi_1}(s_t)=0.05 - \min(\lvert\theta_t\rvert,\lvert\omega_t\rvert)$, $\rho_{\psi_2}(s_t)=0.05 - \min(\lvert1.57-\theta_t\rvert,\lvert\omega_t\rvert)$, $\rho_{\psi_2}(s_t)=0.05 - \min(\lvert-1.57-\theta_t\rvert,\lvert\omega_t\rvert)$, $l_1=0.0103$ $l_2=0.0138$ $l_3=0.0083$, $\gamma_{0,i}=\pi$, $\gamma_{\infty,i}=0.01$, $\rho_{max,i}=0,05$, for all $i\in\{1,2,3\}$. Then, we trained the RL agent using Algorithm \ref{alg:1} to obtain policy enforcing desired STL specifications. Figure \ref{fig:3} shows the evolution of robustness values, angle and angular velocity of the pendulum over time. One can readily observe that the robustness values are always inside the constructed funnels, and the property is satisfied. Notice that the dynamics is under-actuated, so one can not use results in \cite{6}. 

\subsection{Mobile Robot Navigation}\label{sc:4.2}
For the second case study, we consider the differential drive mobile robot described by: $x_{t+1} = x_t +\tau v_t\cos{\theta_t},\; {y}_{t+1} = y_t + \tau v_t\sin{\theta_t}, \;{\theta}_{t+1} = \theta_t + \tau \omega_t$, where, $x$ and $y$ represent the location of robot in $x-y$ plane, $\theta$ represents orientation, and $\tau=0.01$ is the sampling time. The actions $v_t\in\{-5,-4.5,\ldots,4.5,5\}$ and $\omega_t\in\{-3,-2.5,\ldots,2.5,3\}$ represent forward and angular velocity, respectively. The STL specification considered is  
\begin{equation*}
\begin{split}
\Phi= &G_{[900,1300]}\; \big(\|(x,y) - (25,25)\|_2 \leq 2\big)\\ &\land G_{[1600,2000]}\; \big(\|(x,y) - (30,30)\|_2 \leq 2\big). 
\end{split}
\end{equation*}
To satisfy the above STL specification, the agent has to reach inside a circle of radius two centered at (25,25) in the interval from 900 to 1300 timesteps. Further, it should move inside the circle of radius two centered at (30,30). By constructing funnel-based reward, we learn the time-dependent policy using Algorithm \ref{alg:1} to enforce the STL specification. Figure \ref{fig:4} shows the plot of robustness values following the constructed funnel (left plot) and trajectory followed by the trained RL agent in the 2-d plane (middle plot). 

To show the importance of a funnel-based reward structure, we modified the reward function by eliminating the funnel part as described below:
\begin{equation}\label{eq:18}
 r'(s_t,a_t,t) = 
 {
 \begin{cases}
\rho_{\psi_1}    & \text{for } 0\leq t \leq 1300\\
\rho_{\psi_2}    & \text{for } 1300 \leq t \leq 2000.\\
 \end{cases}
 }
\end{equation}
Our ablation study reveals that using robustness values without funnel function $\gamma(t)$ does not help the agent learn the task. The trajectory in the 2-d plane obtained after using the reward function given in \eqref{eq:18} is shown in Figure \ref{fig:4} (right).  

\begin{figure}[t]
    \centering
    \vspace{-0.8em}
    \includegraphics[scale=0.45]{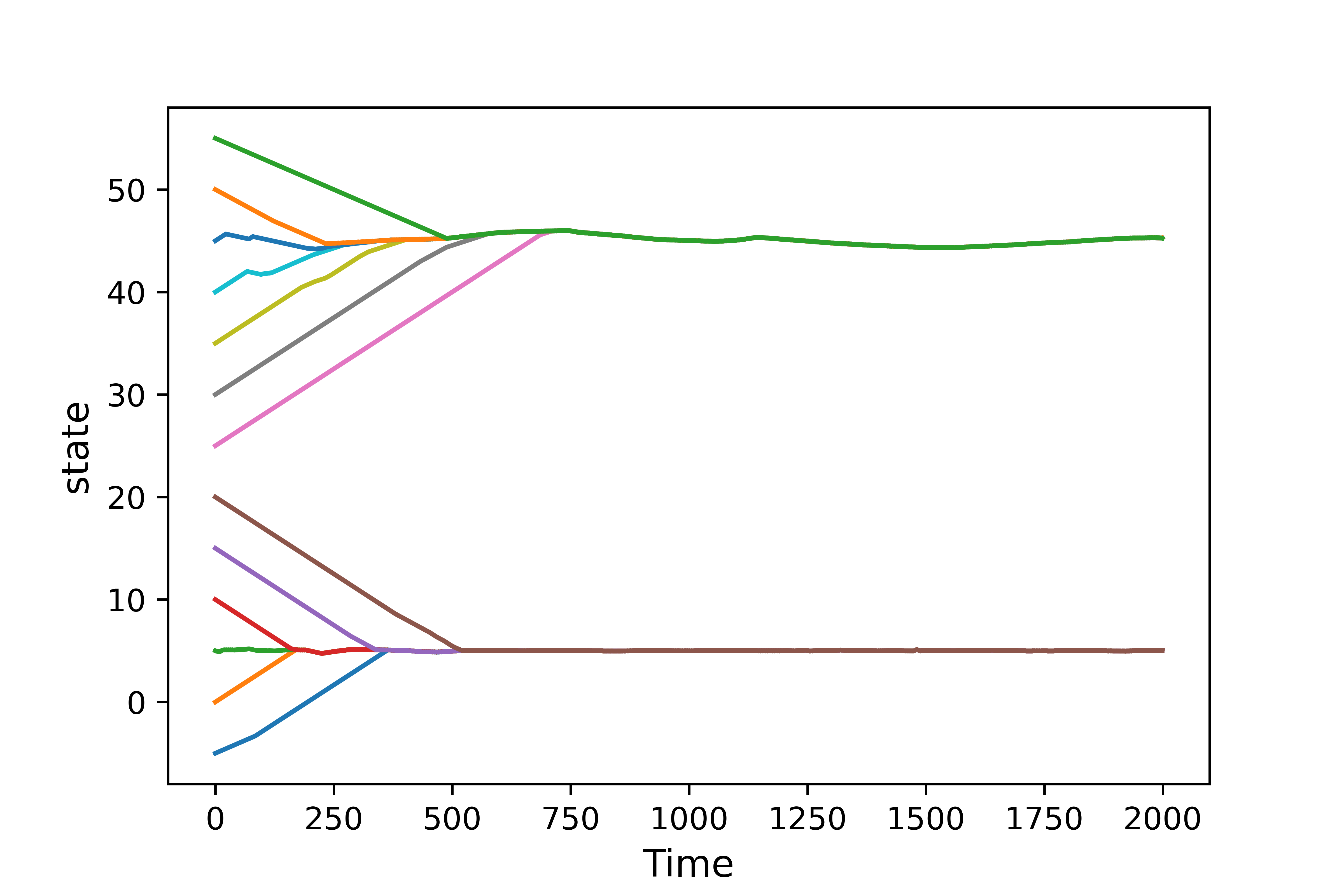}
    \vspace{-0.8em}
\caption{The evolution of trajectories from different initial locations for the discrete-time integrator.}
\label{fig:5}
\end{figure}

\begin{figure}
    \centering
    \includegraphics[scale=0.5]{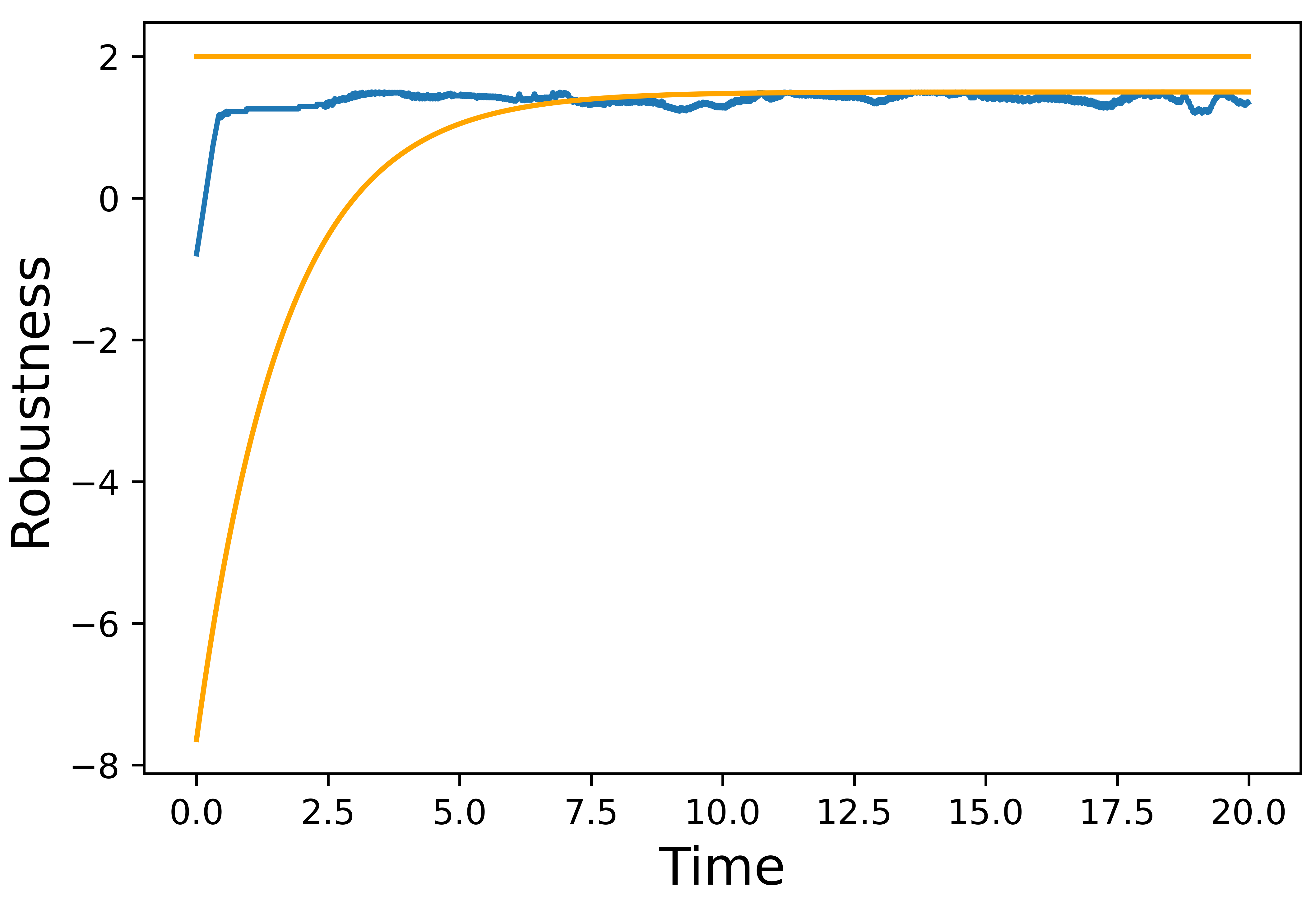}
    \vspace{0.0em}
    \includegraphics[scale=0.5]{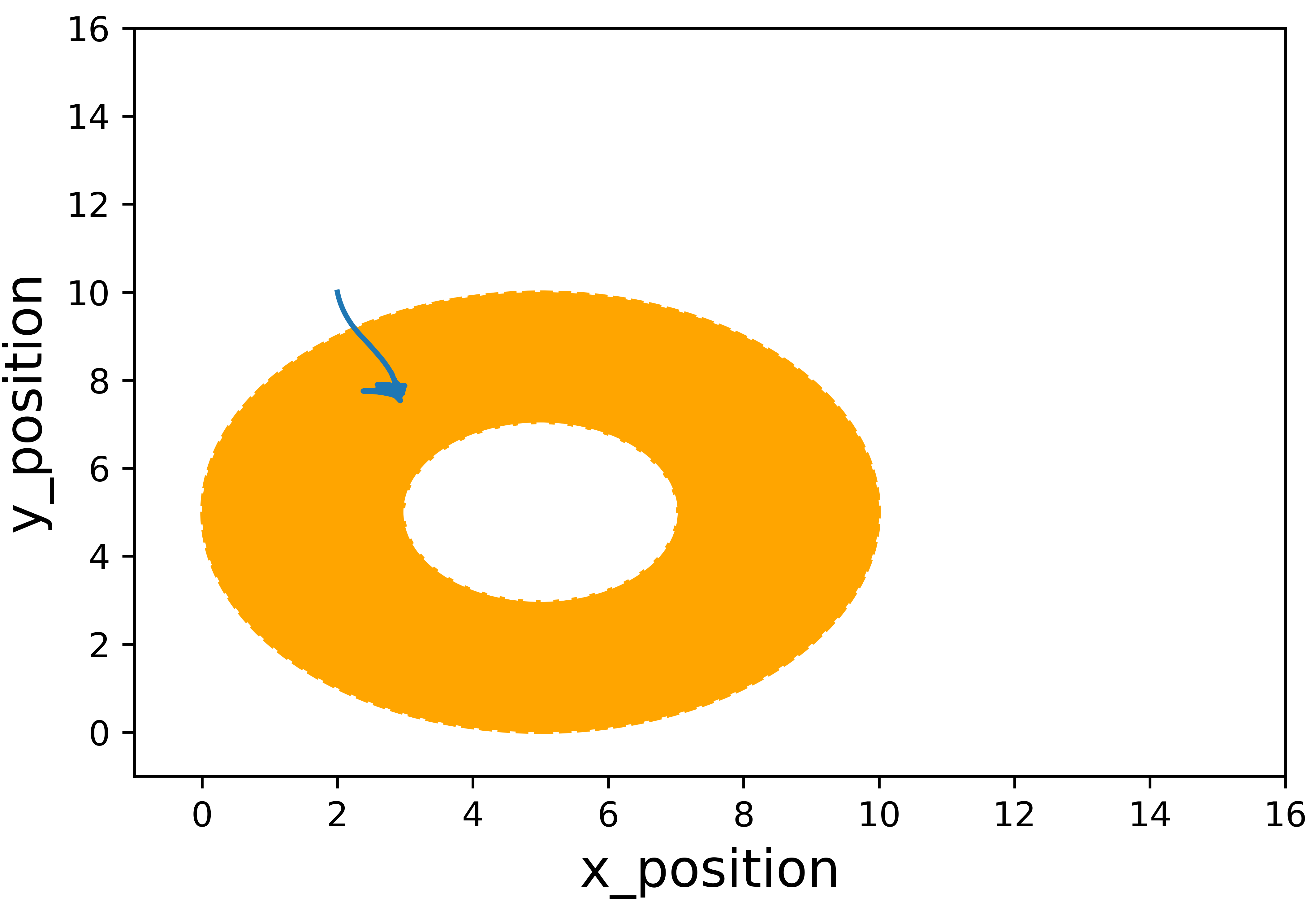}
    \vspace{-0.8em}
\caption{The evolution of robustness values (top), the trajectory of the agent in the 2-d plane (bottom), colored area represents the safe region.}
\label{fig:6}
\end{figure}

\textbf{Specification with convex predicates:} To showcase the applicability of the results for the convex predicate (which is one of the limitations in \cite{6}), we consider the following STL specification: 
\begin{equation}\label{eq:19}
\begin{split}
\Phi= & G_{[300,2000]}\underbrace{\big( \|(x,y) - (5,5)\|_2 \geq 2 }_{\psi_1:(\text{convex predicate})}  \\&\land \underbrace{ \|(x,y) - (5,5)\|_2 \leq 5 \big)}_{\psi_2:(\text{concave predicate})}.
\end{split}
\end{equation}
According to specification \eqref{eq:19}, the robot has to stay inside an annular region centered at (5,5) with the inner radius two and the outer radius 5 for the interval of 300 to 2000 time steps. The agent is always reset at a randomly chosen point in a [0,15]$\times$[0,15] grid. Figure \ref{fig:6} shows the plot of robustness values and the trajectory followed by the robot starting from some random point under the policy learned using our proposed funnel-based STL satisfaction approach in the 2-d plane. One can readily observe that the trajectory achieves the best possible robustness by staying in the centre of the annular region and satisfying the convex predicate $\psi_1$. Hence our method is capable of handling convex predicate. 

\begin{figure}
    \centering
    \includegraphics[scale=0.5]{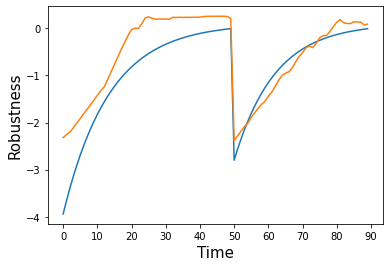}
    \includegraphics[scale=0.5]{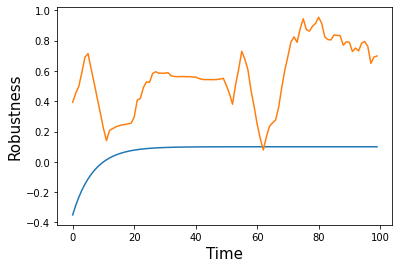}
    \includegraphics[scale=0.5]{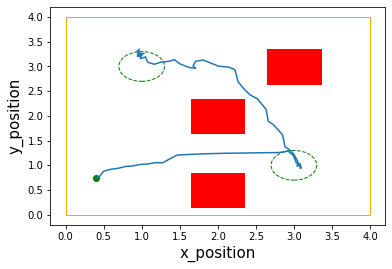}
    \caption{Evolution of robustness value for eventually reaching goals (top), robustness for always avoiding obstacles (middle) and trajectory followed by differential drive robot hardware(bottom)}
    \label{fig:9}
\end{figure}

\textbf{Specification with overlapping time intervals}: We tried STL specification with overlapping time intervals of temporal operator for the differential drive mobile robot and found that our method is able to satisfy specifications with overlapping time intervals as well. The STL specification $\Phi = G_{[0,100]}\; \Big(\big(\|(x,y)-(2,2)\|_{\infty} \leq 2 \big) \land
      \big(\|(x,y)-(2,2)\|_{\infty} \geq 0.5 \big)\land \big(\|(x,y)-(2,0.5)\|_{\infty} \geq 0.5 \big) \land \big(\|(x,y)-(3,3)\|_{\infty} \geq 0.5 \big) \Big) \land F_{[0,50]}\; \big(\|(x,y)-(3,1)\|_2 \leq 0.3 \big) \land F_{[50,90]}\; \big(\|(x,y)-(1,3)\|_2 \leq 0.3 \big) $, and the time intervals are provided in seconds. The robustness plots obtained are given in Figure \ref{fig:9}. We implemented the trained RL agent on hardware, and a video demonstration\footnote[1]{\label{fn:1} The video can be found at this YouTube link: \url{https://www.youtube.com/watch?v=f60-LhD-8PM}} is also available for the same.  

\subsection{Discrete-time Integrator for specification with Disjunction inside Temporal Operator}\label{sc:4.3}
In this section, we show the applicability of our approach to learning policy for STL specification with the disjunction between predicates inside the temporal operator (which is one of the limitations in \cite{6}). Consider the system (discrete-time integrator) with dynamics $x_{t+1} = x_t + \tau v_t$, where $x_t$ is the location at time $t$ and $v_t\in\{-3,-2.5,\ldots,2.5,3\}$ is the velocity given as input to the system. We train the RL agent for the STL specification $G_{[0,2000]}(\varphi_1 \lor \varphi_2)$. Here, $\varphi_1 := \lvert x-5\rvert \leq 5$ and $\varphi_2 := \lvert x-45\rvert \leq 5$. We plotted trajectories of the system in Figure \ref{fig:5} for different initial locations and found that the agent has learnt to reach either $x=5$ or $x=45$. 

\subsection{7-DoF Fetch Mobile Manipulator} 
\begin{figure}
    \centering
    \includegraphics[scale=0.4]{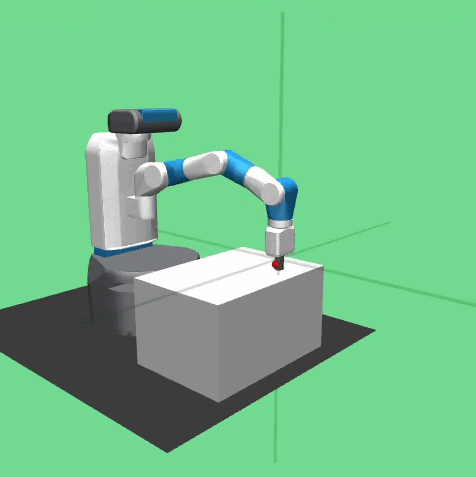}
    \caption{Manipulator arm from the Gymnasium-Robotics suite of tasks}
    \label{fig:8}
\end{figure}

\begin{figure}
    \centering
    \includegraphics[scale=0.5]{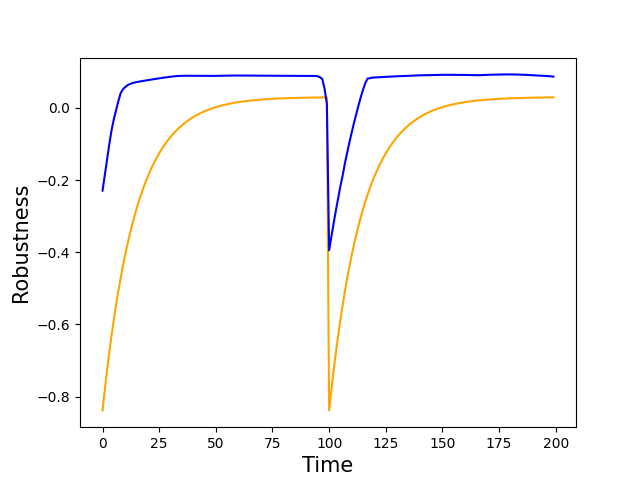}
    \caption{Evolution of robustness values for manipulator arm}
    \label{fig:10}
\end{figure}

Here we consider a manipulator arm (Figure \ref{fig:8}) environment \cite{gymnasium} of "Gymnasium-Robotics" suite of tasks. The task of the manipulator is to move to different points in 3D space according to different time intervals. The actual task is defined using the STL specification $     \Phi = G_{[50,100]}\; \big(\|(x,y,z)-(1.5,0.43,0.47)\|_2 \leq 0.1\big) 
     \land G_{[150,200]}\; \big(\|(x,y,z)-(1.5,1.05,0.47)\|_2 \leq 0.1\big)
$. Because of the continuous action space, we used the TD3 algorithm \cite{fujimoto2018addressing} to train the RL agent. The robustness plot is given as Figure \ref{fig:10} and a video demonstration\footref{fn:1} is also available for the manipulator. The robustness plot shows that our method is capable of handling both continuous state and action space. To substantiate our claim we will provide robustness results on two tasks with continuous action space.

\subsection{DeepMind Cartpole Balance}
\begin{figure}
    \centering
    \includegraphics[scale=0.5]{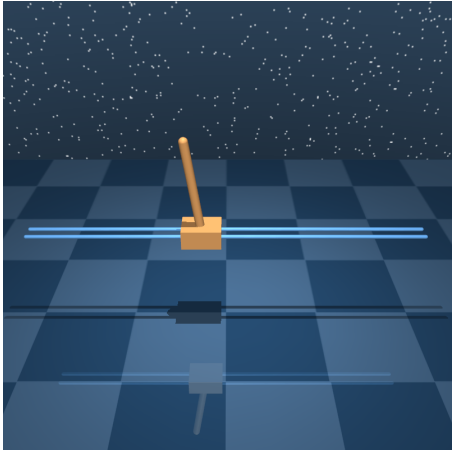}
    \includegraphics[scale=0.5]{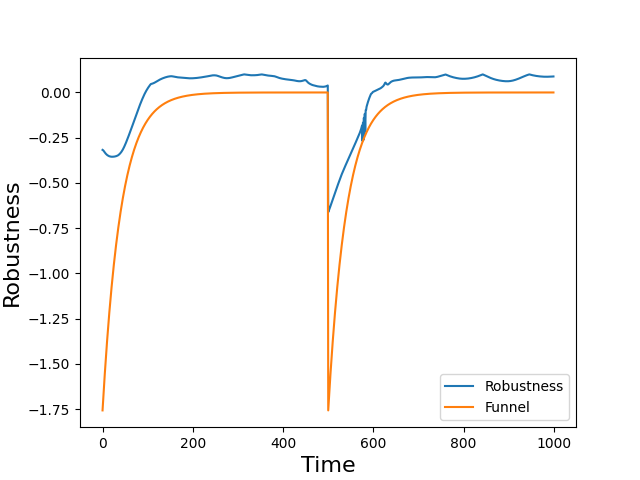}
    \caption{DeepMind control suite CartPole-Balance task(top). Evolution of robustness values for cartpole (bottom).}
    \label{fig:11}
\end{figure}

We test our proposed method using the CartPole-Balance benchmarking task from the DeepMind control suite (\cite{tassa2018deepmind}). In the task considered the RL controller has to balance the pole at one position on the $x$-axis during a particular time interval, and subsequently, the pole has to be balanced at different positions. The temporal constraints-based task is described using the STL specification: $\Phi = G_{[400,500]}\Big(|x-0.5|\leq 0.2 \land |\theta|\leq 0.1 \land |\omega| \leq 0.5\Big) \land G_{[900,1000]}\Big(|x+0.5|\leq 0.2 \land |\theta|\leq 0.1 \land |\omega| \leq 0.5\Big)$. Here $x$ is the position of the cart on the x-axis, $\theta$ is the angle made by the pole with position $y$-axis, and $\omega$ is the angular velocity. The robustness plot is shown in Figure \ref{fig:11}. One can readily observe that the robustness of learned policy satisfies the funnel constraints which ensures satisfaction of specifications.     

\subsection{DeepMind Ball-in-Cup}

\begin{figure}
    \centering
    \includegraphics[scale=0.5]{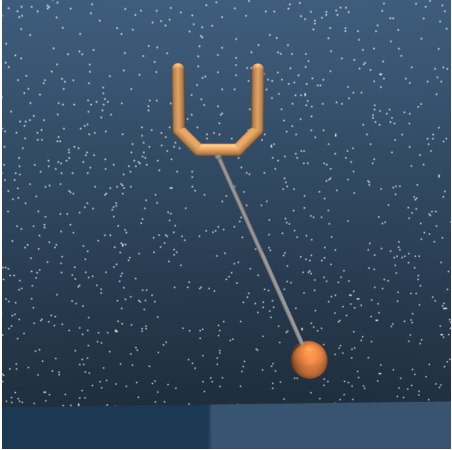}
    \includegraphics[scale=0.5]{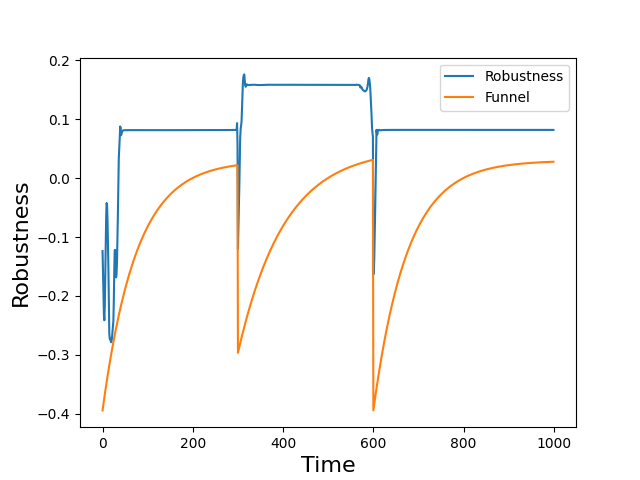}
    \caption{DeepMind control suite Ball-in-cup task(top). Evolution of robustness values for Ball-in-cup task (bottom).}
    \label{fig:12}
\end{figure}

We further evaluated our method on the ball-in-cup benchmarking task from the DeepMind control suite (\cite{tassa2018deepmind}). In this task, the RL agent has to move the cup to capture the ball first, then release the ball, and finally capture the ball again with specific time constraints. The exact task is described by the STL specification as: $ \Phi = G_{[200,300]}\Big(\|(x,z)_{ball} -(x,z)_{cup}\|_2 \leq 0.1\Big) \land G_{[500,600]}\Big(\|(x,z)_{ball} -(x,z)_{cup}\|_2 \geq 0.2\Big) \land G_{[800,1000]}\Big(\|(x,z)_{ball} -(x,z)_{cup}\|_2 \leq 0.1\Big) $. Here, $(x,z)_{ball}$ and $(x,z)_{cup}$ are the location of the ball and the cup respectively in a 2-D plane. The robustness plot is available in Figure \ref{fig:12} which implies satisfaction of specification.

\subsection{Comparative Study}
\begin{figure}[H]
    \centering
    \includegraphics[scale=0.26]{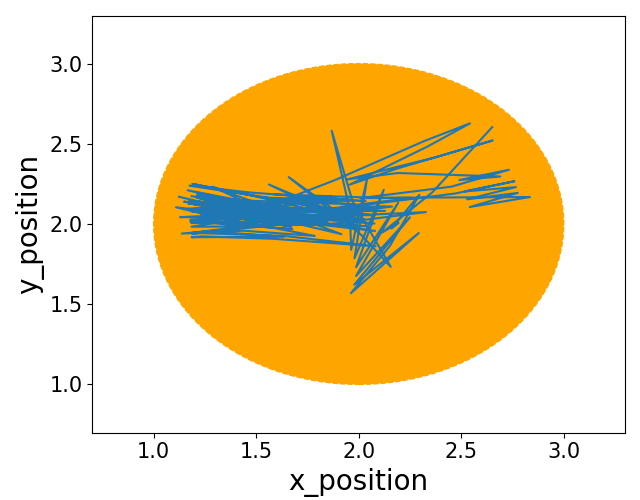}
    \includegraphics[scale=0.26]{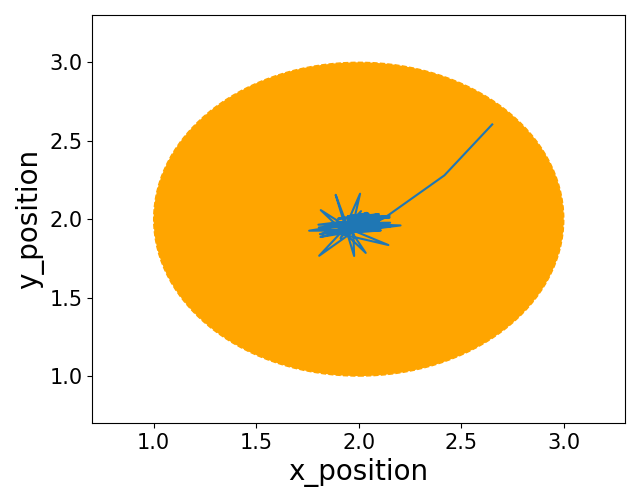}
    \vspace{-0.8em}
\caption{Trajectories generated using the flag-based method \cite{4} (left) and the proposed method (right).}
\label{fig:7}
\end{figure}

We compare our proposed method for STL satisfaction with the flag-based method proposed in \cite{4} by training the RL agent for the differential drive mobile robot (Section \ref{sc:4.2}) considering the STL specification $\Phi = G_{[0,200]}\; \big(\|(x,y)-(2,2)\|_2 \leq 1\big)$. 
We calculated the robustness value by taking a minimum of robustness for all time steps and obtained a robustness value of 0.938 for our method and a robustness value of 0.102 for the method suggested in \cite{4}. Also, from Figure \ref{fig:7}, it is clear that the trajectory generated by using our approach appears more robust. Further, we obtain an execution time of 237.72 minutes for our proposed method as compared to 236.84 minutes for \cite{4} that shows our method achieves better robustness while consuming almost the same amount of time as the state-of-the-art method in \cite{4}. We could not compare our method on infinite state space tasks with the method proposed in \cite{3} because the method is proposed for finite state space only.    

\section{Conclusion}\label{conclusion}
In this paper, we proposed a tractable method to learn a controller for robust satisfaction of STL specification using time-aware deep Q-learning. We described how funnel functions could be used to design reward functions for reinforcement learning algorithms that allow learning controllers for STL specifications. We showed the performance of our method on various environments, such as the pendulum and mobile robot, in accomplishing time-constraint sequential goals. One of the significant advantages of the proposed approach is the satisfaction of the STL formulas with convex predicates. Further, we demonstrated, using a simple environment, that we can learn the controller for STL formula with disjunction operator inside temporal operator. 

\bibliography{reference}
\bibliographystyle{IEEEtran}

\end{document}